\documentclass[twocolumn,prl,showpacs]{revtex4}%
\usepackage{graphicx}%
\usepackage{amsmath}%
\usepackage{amsfonts}%
\usepackage{amssymb}
\usepackage{bm}

\def\s{{\sigma}}
\def\e{{\epsilon}}
\def\k{{ {\bm k} }}

\def\q{{ {\bm q} }}
\def\Q{{ {\bm Q} }}

\def\w{{\omega}}
\def\a{{\alpha}}

\allowdisplaybreaks[4]

\begin{document}
\title{
Non-Fermi-Liquid-Like Behaviors and Superconductivity
Driven by Orbital Fluctuations in Iron Pnictides:
Analysis by Fluctuation-Exchange Approximation
}
\author{Seiichiro \textsc{Onari}$^{1}$,
and Hiroshi \textsc{Kontani}$^{2}$}
\date{\today }

\begin{abstract}
We study the five-orbital Hubbard-Holstein model
for iron pnictides with small electron-phonon interaction
due to Fe-ion Einstein oscillators.
Using the fluctuation-exchange (FLEX) approximation, 
orbital fluctuations 
evolve inversely proportional to the temperature,
and therefore the resistivity shows linear or convex $T$-dependence 
for wide range of temperatures.
We also analyze the Eliashberg gap equation, and show that 
$s$-wave superconducting state without 
sign reversal ($s_{++}$-wave state) emerges
when the orbital fluctuations dominate the spin fluctuations.
When both fluctuations are comparable, their competition gives
rise to a nodal $s$-wave state.
The present study offers us a unified explanation
for both the normal and superconducting states.
\end{abstract}

\address{
$^1$ Department of Applied Physics, Nagoya University and JST, TRIP, 
Furo-cho, Nagoya 464-8602, Japan. 
\\
$^2$ Department of Physics, Nagoya University and JST, TRIP, 
Furo-cho, Nagoya 464-8602, Japan. 
}
 

\sloppy

\maketitle

\begin{figure}[!htb]
\includegraphics[width=0.9\linewidth]{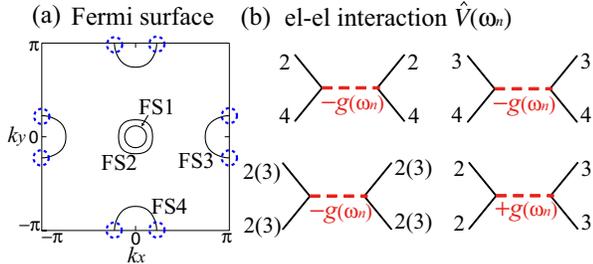}
\caption{(color online) (a) FSs in the unfolded zone.
The dotted circles represent the cold-spot given by the 
orbital fluctuation theory.
The cold-spot is composed of $xz/yz$-orbitals.
(b) Phonon-mediated el-el interaction ($\hat{V}$) for $2,3,4$ orbitals.}
\label{fig1-4}
\end{figure}

The many-body electronic states and the pairing mechanism 
in iron pnictides have been significant open problems.
By taking account of the Coulomb interaction 
and the nesting of the Fermi surfaces (FSs) in Fig.\ref{fig1-4} (a),
fully-gapped sign-reversing $s$-wave state ($s_\pm$-wave state) 
had been proposed 
 \cite{Kuroki,Mazin}.
Experimentally,
both $T_{\rm c}$ and antiferro (AF) spin correlation
increases as $x$ decreases in BaFe$_2$(As$_{1-x}$P$_x$)$_2$ \cite{Nakai}.
In contrast, $T_{\rm c}$ in LaFeAsO$_{1-x}$F$_x$ at $x=0.14$ 
increases from 23 K to 43 K by applying the pressure, whereas
AF spin correlation is almost unchanged \cite{Fujiwara}.
Thus, the relationship between $T_{\rm c}$ and 
spin fluctuation strength seems to depend on compounds.

On the other hand, orbital-fluctuation-mediated $s$-wave state 
without sign reversal ($s_{++}$-wave state) had been proposed 
based on the five-orbital Hubbard-Holstein (HH) model 
\cite{Kontani-Onari,Saito},
which is the Hubbard model introduced in Ref. \cite{Kuroki} 
with the addition of the electron-phonon ($e$-ph) interaction term 
due to Fe-ion Einstein oscillations.
Within the random-phase-approximation (RPA),
it was found that 
$d$-orbital fluctuation is induced by small $e$-ph interaction, 
not by the Coulomb interaction alone.
Especially, empirical relation between $T_{\rm c}$ and the As-Fe-As bond angle
(Lee plot) \cite{Lee} is naturally explained.
Recently, theoretically predicated orbital fluctuations 
\cite{Kontani-Onari,Saito}
had been detected via the substantial softening of shear modulus
 \cite{Yoshizawa}.
The $s_{++}$-wave state is consistent with the robustness of $T_{\rm c}$
against randomness \cite{Onari-impurity,Sato-imp,Nakajima} 
as well as the ``resonance-like'' peak structure 
in the neutron inelastic scattering \cite{Onari-resonance}.

However, spin/orbital fluctuations obtained by the RPA are
reduced by the self-energy correction.
Therefore, to confirm the orbital fluctuation scenario,
it is desired to analyze the many-body electronic states beyond the RPA.
For this purpose, the fluctuation-exchange (FLEX) approximation \cite{Bickers}
would be appropriate in that the absence of spin/orbital order in 2D systems, 
known as Mermin-Wagner theorem, is rigorously satisfied \cite{Mermin-Wagner}.

In this letter, we analyze the five-orbital HH model
for iron pnictides using the FLEX approximation.
In the normal state, large orbital fluctuations induce
highly anisotropic quasiparticle lifetime 
on the FSs as well as 
the $T$-linear or $T$-convex resistivity $\rho$
 \cite{Hall,Kasahara-RH,Eisaki}.
The large orbital fluctuations also introduce
the $s_{++}$-wave superconducting (SC) state for wide range of parameters,
and the competition between orbital and spin fluctuations
lead to the nodal $s$-wave state.
We propose that the orbital fluctuation is the origin of 
both the $s_{++}$-wave SC state and the
non-Fermi-liquid-like behavior in the normal state.

In the FLEX approximation \cite{Bickers},
the $5\times5$ self-energy matrix $\hat{\Sigma}$
in the orbital representation is given by
\begin{equation}
\Sigma_{l_1l_3}(k)=\frac{T}{N}\sum_q\sum_{l_2l_4}V_{l_1l_2,l_3l_4}^\Sigma(q)G_{l_2l_4}(k-q),
\label{eqn:Sigma}
\end{equation}
where $N$ is the number of $\bm{k}$ meshes,
and we denote $k=(\bm{k},\epsilon_n)$ with fermion Matsubara frequency
$\epsilon_n=(2n+1)\pi T$, and $q=(\bm{q},\omega_n)$ with $\omega_n=2n\pi T$.
Here, $l_i$ represents the $Z^2$, $XZ$, $YZ$, $X^2-Y^2$ and $XY$ orbitals,
which are hereafter denoted as 1, 2, 3, 4 and 5, respectively \cite{Kuroki}.
$\hat{G}$ is the $5\times 5$ Green function matrix in the orbital basis,
and $\hat{V}^\Sigma$ is the $25\times 25$ interaction term
for the self-energy given as \cite{Takimoto}
\begin{eqnarray}
\hat{V}^\Sigma(q)&=&\frac{3}{2}\hat{\Gamma}^s\hat{\chi}^s(q)\hat{\Gamma}^s+\frac{1}{2}\hat{\Gamma}^c\hat{\chi}^c(q)\hat{\Gamma}^c \nonumber\\
&-&\frac{1}{4}(\hat{\Gamma}^s-\hat{\Gamma}^c)\hat{\chi}^{\rm irr}(q)(\hat{\Gamma}^s-\hat{\Gamma}^c)+\frac{3}{2}\hat{\Gamma}^s+\frac{1}{2}\hat{\Gamma}^c,
\label{eff}
\end{eqnarray}
where the irreducible susceptibility is given by
\begin{equation}
\chi^{\rm irr}_{l_1l_2,l_3l_4}(q)=-\frac{T}{N}\sum_kG_{l_1l_3}(k+q)G_{l_4l_2}(k),
\end{equation}
and the spin (orbital) susceptibility is obtained as
$\hat{\chi}^{s(c)}=\hat{\chi}^{\rm irr}(1-\hat{\Gamma}^{s(c)}\hat{\chi}^{\rm irr})^{-1}$.
Here, $\hat{\Gamma}^{s}=\hat{S}$ ($\hat{\Gamma}^{c}=-\hat{C}-2\hat{V}(\w_n)$)
is the irreducible vertex for spin (charge) channel;
${\hat S}$ and ${\hat C}$ represent the Coulomb interaction 
in the multiorbital model introduced in Refs. 
 \cite{Kuroki,Takimoto,Kontani-Onari,Saito};
Their matrix elements consist of
the intra-orbital Coulomb $U$, the inter-orbital Coulomb $U'$, Hund's
coupling $J$ and the pair hopping $J'$.
Here we assume that $J=J'$ and $U=U'+2J$.

The vertex $\hat{V}(\w_n)$ in $\hat{\Gamma}^{c}$
represents the electron-electron (el-el) 
interaction mediated by $e$-ph interaction.
For example, we show non-zero $V_{ll',mm'}$ for $l,l',m,m'=2,3,4$
in Fig. \ref{fig1-4} (b),
where $g(\omega_n)=g\omega_{\rm D}^2/(\omega_n^2+\omega_{\rm D}^2)$
is proportional to the phonon Green function;
$g=g(0)$ is the effective el-el interaction for $\w_n=0$, and 
$\omega_{\rm D}$ is the Debye frequency \cite{Kontani-Onari}.
Other than Fig. \ref{fig1-4} (b),
$\hat{V}$ has many non-zero off-diagonal elements
as explained in Ref. \cite{Saito},
since the Fe-ion oscillation (non-A$_{1g}$ mode)
induces various inter-orbital transitions.
This fact gives rise to the prominent orbital fluctuations
at low frequencies, while the charge susceptibility 
$\chi^c(\bm{q})=\sum_{lm}\chi^c_{ll,mm}(\bm{q})$ is not enhanced 
due to the cancellation \cite{Kontani-Onari,Saito}.
In the present study, we drop ladder-type diagrams by $g(\w_n)$, 
which is justified when $\w_{\rm D}\ll E_{\rm F}$ \cite{Kontani-Onari,Saito}.
For the same reason, $g(\w_n)$ is absent in ${\hat \Gamma}^s$.


In the FLEX approximation,
we obtain $\hat{G}$ and $\hat{\Sigma}$ self-consistently
using the Dyson equation $\hat{G}^{-1}=(\hat{G}^0)^{-1}-\hat{\Sigma}$.
In multiband systems, the FSs are modified from the original FSs
due to the self-energy correction.
To escape from this difficulty, we subtract the constant term
$[\hat{\Sigma}(\bm{k},+i0)+\hat{\Sigma}(\bm{k},-i0)]/2$
from the original self-energy, corresponding to the
elimination of double-counting terms between LDA and FLEX
\cite{Ikeda}.
Hereafter, 
we fix $J/U=1/6$, $\omega_{\rm D}=0.02$eV, and the electron filling
$n=6.1$ except for Fig. \ref{fig3}.
Because of the smallness of the FSs in Fig. \ref{fig1-4},
fine $\bm{k}$ meshes are required for a quantitative study.
We take $N=128\times128$ $\bm{k}$ meshes 
that is four times of that used in Ref. \cite{Ikeda},
and 1024 Matsubara frequencies.
Then, we obtain reliable numerical results for $T\ge0.01$eV.

\begin{figure}[!htb]
\includegraphics[width=\linewidth]{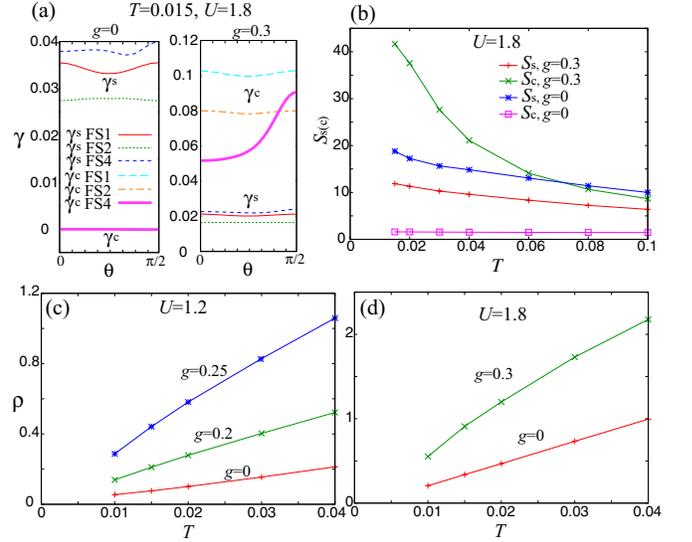}
\caption{(color online) 
(a) $\k$-dependence of $\gamma^{s(c)}$ induced by the spin (orbital)
 fluctuations on each FS.
(b) $T$-dependence of $S_{s(c)}=(1-\a_{s(c)})^{-1}$. 
(c),(d) $T$-dependence of $\rho$.
$\rho=1$ corresponds to $(\hbar a_c)/e^2\sim 300\mu$cm$\Omega$
when the interlayer distance is $a_c=0.6$nm.
}
\label{fig1}
\end{figure}

We begin with the electronic property in the normal state.
Hereafter, the unit of energy is eV.
First, we discuss the quasiparticle damping rate
$\gamma_{\bm k}$ on each FS, which is given by the imaginary part 
of the self-energy in the band-diagonal representation.
In Fig. \ref{fig1} (a), 
$\gamma^{s(c)}_{\bm{k}}$ represents the damping due to 
spin (orbital) fluctuations for $T=0.015$ and $U=1.8$,
which is given by substituting
$\hat{V}^\Sigma=\frac{3}{2}\hat{\Gamma}^s\hat{\chi}^s\hat{\Gamma}^s$
$(\frac{1}{2}\hat{\Gamma}^c\hat{\chi}^c\hat{\Gamma}^c)$ 
in eq. (\ref{eqn:Sigma}).
The horizonal axis is the azimuth angle for $\bm{k}$ point 
with the origin at $\mathrm{\Gamma}$(M) point for FS1,2 (FS4).
The relation $\gamma_\k\approx\gamma^{s}_\k+\gamma^{c}_\k$ is satisfied 
since the third term in eq. (\ref{eff}) is very small.
We will see below that the value $U=1.8$ can reproduce moderate
AF spin fluctuations observed in $e$-doped compounds, and
it is consistent with $U\sim2$ reported by 
x-ray absorption spectroscopy (XAS) \cite{x-ray}.

In the case of $g(0)=g=0$, the relation $\gamma^s\gg\gamma^c$ holds since 
the orbital fluctuation is very small.
However, $\gamma^c$ gradually develops with $g$,
and $\gamma^s\sim\gamma^c$ at $g=0.26$.
In Fig. \ref{fig1} (a), $\gamma^c\gg\gamma^s$ at $g=0.3$;
the corresponding dimensionless coupling is just $\lambda\equiv gN(0)\sim0.2$.
In contrast, $\gamma^s$ decreases with $g$,
due to the suppression of $\hat{\chi}^s$ by $\gamma^c$.

As shown in Fig. \ref{fig1} (a),
the momentum dependence of $\gamma_\k^s$ on each FS is small,
although AF spin correlation is well developed.
In contrast, $\gamma_\k^c$ on FS4 (e-pockets) 
takes the minimum value at $\theta\sim0$ for $g=0.3$;
this ``cold-spot'' is important for the transport phenomena.
Since the cold spot is on the e-pocket,
the Hall coefficient $R_{\rm H}$ will be negative,
which is consistent with experiments
\cite{Sato-RH,Hall,Kasahara-RH}.
In the case of high-$T_{\rm c}$ cuprates, 
various non-Fermi-liquid-like transport phenomena 
(e.g., violation of Kohler's rule) originate from the 
cold/hot spot structure as well as the backflow 
(=current vertex correction) due to spin fluctuations
\cite{ROP}.
Therefore, appearance of the cold spot in Fig. \ref{fig1} (a)
indicates that the orbital fluctuations are the origin of 
striking non-Fermi-liquid-like transport phenomena in iron pnictides
\cite{Sato-RH,Hall,Kasahara-RH}.

In Fig. \ref{fig1} (b), we show how the orbital and spin fluctuations 
develop as $T$ decreases:
In the FLEX, the spin (orbital) susceptibility is enhanced by the 
spin (orbital) Stoner enhancement factor $S_{s(c)}=(1-\alpha_{s(c)})^{-1}$, 
where $\a_{s(c)}$ is the maximum of the largest eigenvalue of 
$\hat{\Gamma}^{s(c)}\hat{\chi}^{\rm irr}(\bm{q},0)$ with respect to $\bm{q}$.
Then, $\alpha_{s,c}=1$ corresponds to the spin/orbital order, although
it is prohibited in 2D systems by the Mermin-Wagner theorem
 \cite{Mermin-Wagner}.
In case of $U=1.8$ and $g=0$,
large $S_{s}\ (\gtrsim10)$ is induced at $\q\approx\Q\equiv(\pi,0)$
(i.e., $\chi^s(\Q,0)\propto S_{s}$).
$S_{s}$ gradually increases as $T$ drops, which is a typical critical 
behavior near the AF magnetic quantum-critical-point (QCP)
\cite{Moriya}.
When $g>0$, $\chi^c(\bm{q},0)$ is enhanced 
at $\q=\bm{0}$ and $\q=\Q$ almost equivalently \cite{Saito}.
At $g=0.3$, large $S_{c}\ (\gg10)$ is induced at 
both $\q\approx\Q$ and ${\bm 0}$,
and it increases approximately proportional to $T^{-1}$.
Thus, it is confirmed that both ferro- and AF-orbital fluctuations 
show critical evolusions near the orbital QCP.

Next, we discuss the resistivity $\rho$ due to 
the orbital and spin fluctuations.
By neglecting the backflow, the conductivity is obtained by
\begin{eqnarray}
\sigma_{xx}=\frac1N \sum_{\bm{k},\alpha}\int_{-\infty}^{\infty}
 \frac{d\omega}{\pi}\left(-\frac{\partial f(\omega)}{\partial \omega}\right)
 \left|v^x_{\a,\k} G_{\k,\a}(\w+i0)\right|^2 ,
\end{eqnarray}
where $\a$ is the band index,
$f(\omega)$ is the Fermi distribution function, 
$v^x_{\a,\k}$ is the velocity of band $\a$,
and $G_{\k,\a}(\w+i0)$ is the retarded Green function 
for band $\a$ in the FLEX approximation.
Figure \ref{fig1} (c) and (d) show the obtained the resistivity $\rho=1/\s_{xx}$
for $U=1.2$ and 1.8:
In case of $U=1.2$, 
$\rho$ shows a conventional sublinear (concave) $T$-dependence at $g=0$.
$\rho$ increases with $g$ due to the orbital fluctuations, and
almost $T$-linear resistivity is realized at $g=0.2$.
At $g=0.25$, $\rho$ shows a superlinear (convex) $T$-dependence.
In case of $U=1.8$,
$\rho$ is linear-in-$T$ at $g=0$, while it 
shows a clear superlinear $T$-dependence at $g=0.3$.
Thus, we stress that
non-Fermi-liquid resistivity is not a direct evidence for 
the spin fluctuations.
In $Ln$FeAsO compounds,
$T_{\rm c}$ increases as the radius of ranthanide ion $Ln^{3+}$ decreases,
and the $T$-dependence of $\rho$ changes from concave
to convex
\cite{Eisaki}.
This experimental correlation between $T_{\rm c}$ and $\rho(T)$
may be understood in terms of the orbital fluctuation scenario.

\begin{figure}[!htb]
\includegraphics[width=\linewidth]{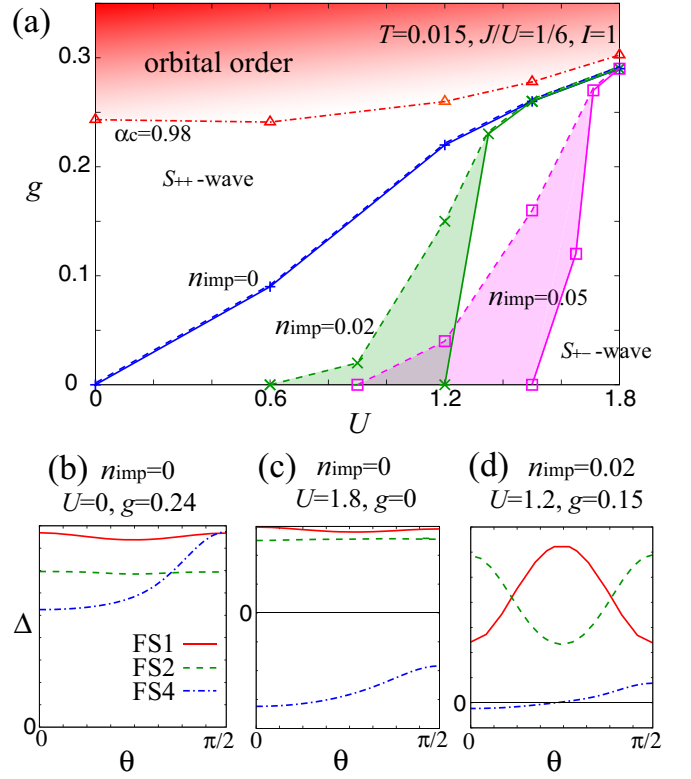}
\caption{(color online) (a) $U$-$g$ phase diagram 
 given by solving the linearized Eliashberg equation at $T=0.015$.
 Nodal $s$-wave gap state is obtained in the shaded area
 for $n_{\rm imp}=0.02$ and 0.05, and
 solid lines (dotted lines) represent the boundary between 
 fully-gapped $s_{+-}$-wave ($s_{++}$-wave) state.
 Dashed-dotted line denotes $\alpha_c=0.98$.
 (b) $s_{++}$-wave gap ($\lambda_E=0.59$)
 for $U=0$ and $g=0.24$.
 (c) $s_{+-}$-wave gap ($\lambda_E=0.49$)
 for $U=1.8$ and $g=0$ and.
 (d) Nodal $s$-wave gap ($\lambda_E=0.28$)
 for $U=1.2$ and $g=0.15$.
}
\label{fig2}

\end{figure}

Next, we discuss the SC state.
In the presence of dilute impurities ($n_{\rm imp}\ll1$),
the linearized Eliashberg equation in the orbital basis is
\cite{Kontani-Onari}:
\begin{eqnarray}
\lambda_{\rm E}\Delta_{ll'}(k)&=-&\frac{T}{N}\sum_{k',m_i}W_{lm_1,m_4l'}(k-k')
G'_{m_1m_2}(k')\nonumber\\
&\times&\Delta_{m_2m_3}(k')
G'_{m_4m_3}(-k')+\delta\Sigma^a_{ll'}(\epsilon_n),
\label{eqn:Eliash}
\end{eqnarray}
where $\Delta_{ll'}(k)$ is the gap function and
$\lambda_{\rm E}$ is the eigenvalue that reaches unity at $T=T_c$.
$\delta{\hat Sigma}^a$ represents the impurity-induced gap function.
$(\hat{G'})^{-1}=(\hat{G})^{-1}-\delta\hat{\Sigma}^n$,
where $G$ is the Green function given by eq. (\ref{eqn:Sigma}),
and $\delta{\hat \Sigma}^n$ is the impurity-induced normal self-energy.
The pairing interaction ${\hat W}$ in eq. (\ref{eqn:Eliash}) is
\begin{equation}
\hat{W}(q)=\frac{3}{2}\hat{\Gamma}^s\hat{\chi}^s(q)\hat{\Gamma}^s-\frac{1}{2}\hat{\Gamma}^c\hat{\chi}^c(q)\hat{\Gamma}^c+\frac{1}{2}\hat{\Gamma}^s-\frac{1}{2}\hat{\Gamma}^c,
\label{eqn:W}
\end{equation}
where $\hat{\chi}^{s,c}$ is given by the FLEX approximation for $n_{\rm imp}=0$, 
because of the fact that the fully self-consistent FLEX with impurity-induced 
self-energy leads to unphysical reduction in $\chi^s$,
unless vertex correction is taken into account \cite{ROP}.
The first (second) term in eq. (\ref{eqn:W}) works to 
set $\Delta_{\rm FS1,2}\cdot \Delta_{\rm FS3,4}<0$ ($>0$).

In the $T$-matrix approximation,
$\delta{\hat \Sigma}^{n,a}$ is given as
\begin{eqnarray}
\delta\Sigma^n_{ij}(\epsilon_n)&=&n_{\rm imp}T_{ij}(\epsilon_n) ,
 \label{eqn:dSn} \\
\delta\Sigma^a_{ij}(\epsilon_n)&=&n_{\rm imp}\sum_{lm}T_{il}(\epsilon_n)
f_{lm}(\epsilon_n)T_{jm}(-\epsilon_n)
 \label{eqn:dSa},
\end{eqnarray}
where 
$T_{ij}(\e_n)\equiv I(1- I{\hat g}(\e_n))^{-1}$
is the $T$-matrix in the normal state \cite{Onari-impurity};
${\hat g}(\e_n)\equiv \frac1N\sum_\k{\hat G}_\k(\e_n)$
is the local normal Green function, and $I$ is the local 
impurity potential that is diagonal in the orbital basis.
We put $I=1$ hereafter.
In eq. (\ref{eqn:dSa}), 
$f_{ij}(\epsilon_n)=\frac1N\sum_{\bm{k},lm}G_{il}(k)\Delta_{lm}(k)G_{jm}(-k)$
is the linearized local anomalous Green function.

In Fig. \ref{fig2} (a), we show the $U$-$g$ phase diagram 
obtained by the FLEX approximation.
The dashed-dotted line represents the condition $\alpha_c=0.98$
at $T=0.015$, corresponding to $g=0.25\sim0.3$.
(In the RPA, the same condition 
is satisfied for $g=0.21\sim0.23$ \cite{Saito}.)
Therefore, prominent orbital fluctuations emerge
for $\lambda=gN(0)\lesssim0.2$
even if the self-energy correction is taken into account.
On the other hand,
$\alpha_s=0.95$ (0.92) for $U=1.8$ and $g=0$ (0.3)
in the FLEX approximation,
although $U_{\rm cr}=1.25$ for $\alpha_s=1$ in the RPA.
Thus, the renormalization in $\a_s$ is rather larger than that in $\a_c$,
because of the difference in the coefficients (in factor 3)
between the first and the second terms in eq. (\ref{eff}).

Next, we solve eq. (\ref{eqn:Eliash}) with high accuracy
using the Lanczos method at $T=0.015$.
Then, $s_{++}$-wave gap function is obtained around the line $\alpha_c=0.98$;
Figure \ref{fig2} (b) shows the $s_{++}$-wave gap 
for $g=0.24$ and $U=0$ ($\lambda_E=0.59$).
On the other hand, $s_{\pm}$-wave gap is obtained when 
$g$ is sufficiently small;
Figure \ref{fig2} (c) shows the $s_{\pm}$-wave gap 
for $U=1.8$ and $g=0$ ($\lambda_E=0.49$).  
When $n_{\rm imp}=0$, the gap function changes 
from (b) to (c) discontinuously on the phase boundary in Fig. \ref{fig2} (a),
as found in Ref. \cite{Saito}.
When $n_{\rm imp}\ge0.02$, however, 
gap function changes smoothly during the crossover.
Then, line-nodes inevitably appear on FS3,4 in the shaded area
in Fig. \ref{fig2} (a);
Figure \ref{fig2} (d) shows the nodal $s$-wave gap 
for $U=1.2$, $g=0.15$ and $n_{\rm imp}=0.02$ ($\lambda_E=0.28$).
Thus, both regions for $s_{++}$-wave and nodal $s$-wave states are extended 
by small amount of impurities, 
although $\lambda_E$ for the latter state is reduced by impurities.
A nodal $s$-wave solution at $n_{\rm imp}=0$ with larger $\lambda_E$
may be obtained by considering a three-dimensional (3D) 
nodal-line structure in 3D tight-binding model \cite{Mazin2}.

Here, we discuss that line nodes originate from
the competition between the orbital and spin fluctuations:
The electrons at $\theta\sim0$ ($\pi/2$) on FS4
is composed of orbital 2,3 (4).
Since the orbital 4 is absent in FS1,2,
the nesting-driven spin correlation between 
the orbital 2,3 on FS1,2 and the orbital 4 on FS3,4 is weak.
(That is, $\chi^s_{24,42}(\Q)\ll\chi^s_{22,22}(\Q)$.)
On the other hand, both $\chi^c_{24,42}(\q)$ and $\chi^c_{22,22}(\q)$
are developed here \cite{Kontani-Onari,Saito}.
Therefore, when orbital and spin fluctuations are comparable,
$\Delta_{\rm FS1,2}\cdot \Delta_{\rm FS4}$ is negative (positive) 
at $\theta\sim0$ ($\pi/2$)
due to the orbital-dependence on the spin correlation.

\begin{figure}[!htb]
\includegraphics[width=0.99\linewidth]{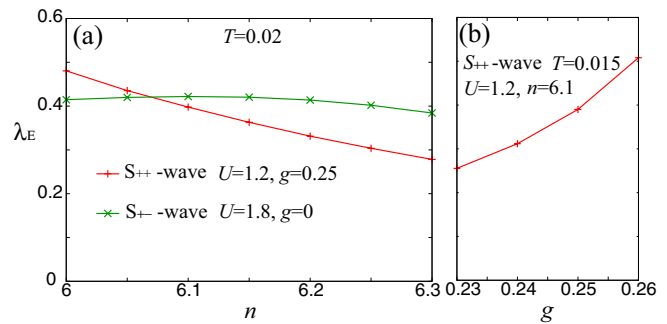}
\caption{(color online) 
(a) $n$ dependence of $\lambda_{\rm E}$ for
 $s_{++}$- and $s_{+-}$-wave states at $T=0.02$ and $n_{\rm imp}=0$. 
(b) $g$ dependence of $\lambda_{\rm E}$ for $s_{++}$-wave state
at $T=0.015$ and $n_{\rm imp}=0$. 
}
\label{fig3}
\end{figure}

In Fig. \ref{fig3} (a), we show the filling dependence of $\lambda_{\rm E}$
for the $s_{++}$-wave state ($U=1.2$, $g=0.25$), and that 
for the $s_{+-}$-wave state ($U=1.8$, $g=0$).
We note that FS1,2 disappear for $n>6.3$. 
The value of $\lambda_{\rm E}$ for the $s_{++}$-wave state decreases 
monotonically with $n$, while $\lambda_{\rm E}$ for the $s_{+-}$-wave state 
is rather insensitive to $n$, 
maybe because the temperature, $T=0.02$, is rather high.
Figure \ref{fig3} (b) shows that $\lambda_{\rm E}$ 
for the $s_{++}$-wave state ($U=1.2$, $n=6.1$)
increases with $g$. 

In summary, we performed the FLEX approximation in the 
multiorbital HH model for iron pnictides.
It was confirmed that orbital-fluctuation-mediated $s_{++}$-wave state
is realized by small $e$-ph interaction $g$.
As increasing the value of $g$, both the $T_{\rm c}$ of $s_{++}$-wave state and 
the resistivity $\rho$ are increased, 
and the latter changes from $T$-concave to $T$-convex.
This correlation between $T_{\rm c}$ and $\rho$ is
consistent with experiment \cite{Eisaki}.
The region of $s_{++}$-wave or nodal $s$-wave states is enlarged
in the presence of small amount of impurities.
Thus, the present orbital fluctuation scenario presents a unified 
explanation for both normal and SC electronic states.

\acknowledgements
We are grateful to M. Sato, Y. Kobayashi, Y. Matsuda, T. Shibauchi,
D.S. Hirashima, Y. Tanaka, K. Yamada, and F.C. Zhang 
for valuable discussions. 
This study has been supported by Grants-in-Aid for Scientific 
Research from MEXT of Japan, and by JST, TRIP.


\end{document}